\documentclass[reprint, groupedaddress, showpacs, showkeys, aps]{revtex4-1}

\usepackage{graphicx} 
\usepackage{dcolumn}  
\usepackage{bm}       

\usepackage[usenames,dvipsnames]{pstricks}
\usepackage{epsfig}
\usepackage{pst-grad} 
\usepackage{pst-plot} 
\usepackage{float}

\usepackage{subfigure}

\usepackage{amsmath}
\usepackage{amssymb}

\newcommand{\R}{\mathbb{R}}
\renewcommand{\phi}{\varphi}
\newcommand{\E}{\mathcal{E}}
\newcommand{\ud}{\mathrm{d}}
\renewcommand{\P}{\mathbb{P}}
\renewcommand{\O}{\mathcal{O}}

\newcommand{\be}{\begin{equation}}
\newcommand{\ee}{\end{equation}}

\begin{document}

\title{Nonlinear waves in networks: a simple approach using the sine--Gordon equation}

\author{Jean-Guy Caputo}
\email{caputo@insa-rouen.fr}
\homepage{http://lmi.insa-rouen.fr/~caputo/}
\affiliation{Laboratoire de Math\'ematiques, INSA de Rouen,\\ 76801 Saint-Etienne du Rouvray, France}

\author{Denys Dutykh}
\email{Denys.Dutykh@univ-savoie.fr}
\homepage{http://www.denys-dutykh.com/}
\affiliation{LAMA, UMR 5127 CNRS, Universit\'e de Savoie, \\ Campus Scientifique, 73376 Le Bourget-du-Lac Cedex, France}

\date{\today}

\begin{abstract}
To study the propagation of nonlinear waves across Y-- and T--type junctions, we consider the 2D sine--Gordon equation as a model and study the dynamics of kinks and breathers in such geometries. The comparison of the energies reveals that the angle of the fork plays no role. Motivated by this, we introduce a 1D effective equation whose solutions agree well with the 2D simulations for kink and breather solutions. For branches of equal width, breather crossing occurs approximately when $v > 1 - \omega$, where $v$ is the breather celerity and $\omega$ is its frequency. We then characterize the breathers in the two upper branches by estimating their velocity and frequency. These new breathers are slower than the initial breather and up-shifted in frequency. In perspective, this study could be generalized to more complex nonlinear waves.
\end{abstract}


\maketitle

\section{Introduction}

\begin{figure}
  \centering
  \includegraphics[width=0.30\textwidth]{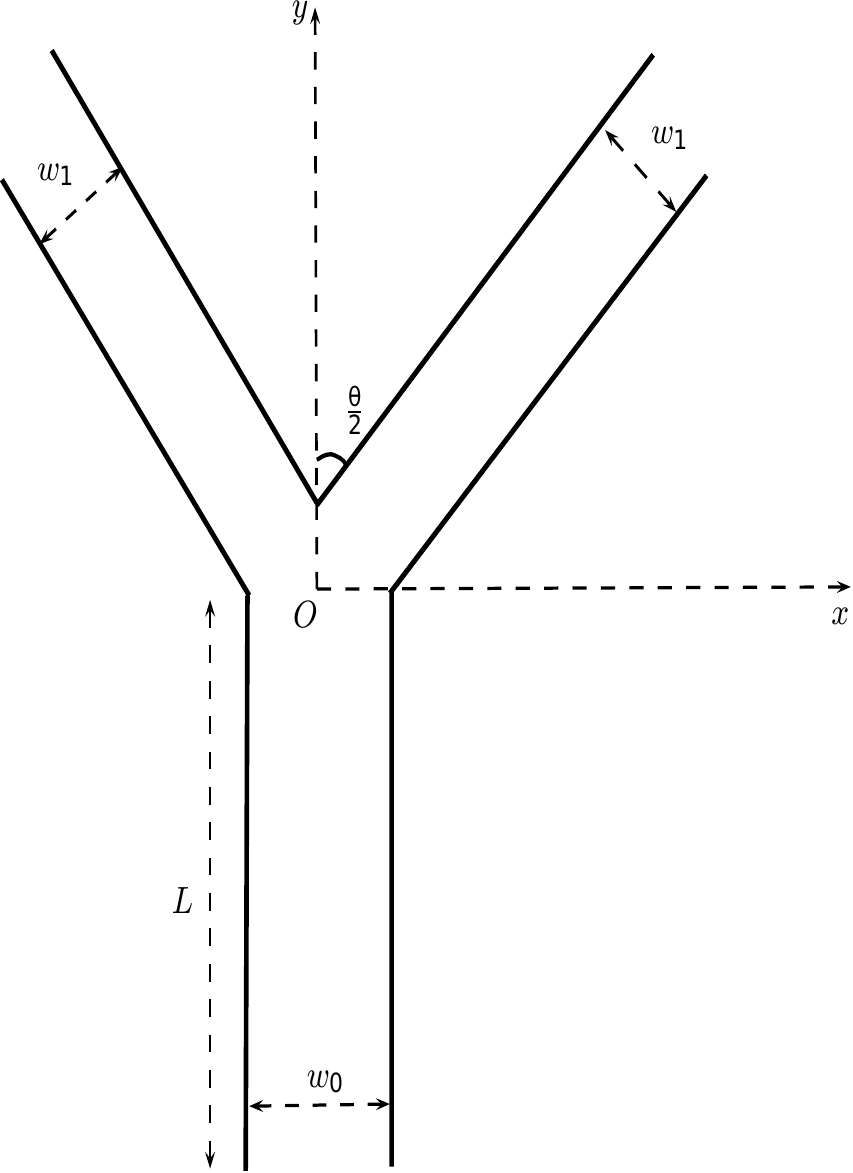}
  \caption{\small\em Sketch of the computational domain $\Omega$ .}
  \label{fig:f1}
\end{figure}

The propagation of nonlinear waves in networks is a very common problem. Examples are the nerve impulse traveling in arrays of neurons, the motion of the pulse wave in the arterial circulatory system or the propagation of waves in the electrical power grid. In general the problem is difficult to tackle because both the equation of motion and the geometry are complex. A first direction is to look at what happens in a simpler geometry like a Y--junction, see Fig.~\ref{fig:f1}. Another simplification is to examine what happens for a linear wave equation. In this context, a number of researchers have examined so-called quantum graphs where the Schr\"odinger equation is solved on a network. See \cite{gs06} for a review. For these linear systems, the scattering formalism can be employed and this gives the reflection and transmission coefficients for a harmonic wave. This is detailed specifically for a Y--junction and for the Klein--Gordon linear wave equation in \cite{bp09}.

In many cases however the nonlinearity cannot be neglected. For fluid systems, one can note the works by Bona \& Cascaval \cite{Bona2008} and Mugnolo \& Rault \cite{Mugnolo2013} who used the Benjamin--Bona--Mahoney (BBM) shallow water equation to describe a fluid network. The authors used the fact that the BBM equation is unidirectional, hence, most of the energy is propagated downstream. Note the numerical work by Nachbin \& Da Silva Simoes \cite{Nachbin2012} where they solved the Boussinesq equations on a junction using a conformal mapping technique. These studies do not provide a simple understanding of the behavior of the waves, in particular one cannot see easily how energy travels across the network.

To address these issues it is easier to consider first a typical nonlinear wave equation. For instance, before tackling the propagation of shallow water waves in a river basin, for which there are two variables, the water elevation and the potential flow, it is useful to consider a simpler nonlinear hyperbolic equation. The sine--Gordon equation is precisely a simpler nonlinear hyperbolic equation that admits localized solutions. Furthermore it is a Hamiltonian system in any dimension, integrable in 1D so that one can test the numerical solutions versus their exact counterpart propagation in a 1D channel. Finally, the sine--Gordon equation arises naturally in the modelling of Josephson junctions as a result of the continuous limit \cite{Bena2002}.

Consider the 2D sine--Gordon equation defined in a Y--junction such as shown in
Fig.~\ref{fig:f1}. A first work on the problem is by Gulevich \& Kusmartsev \cite{Gulevich2006} who examined numerically how kinks propagate in such a system, in the context of Josephson junctions. They showed that the kink needs a sufficient velocity to cross the branch. Here we follow up on this and defined a 2D symmetric junction parameterized by the angle $\theta$ between the branches and their widthes $w_{1,2}$, so that it can go from a Y--junction to a T--junction $\theta=180^\circ$. We solved the 2D problem using the \texttt{FreeFem++} finite element library \cite{Hecht2012}. We made sure that energy was conserved by the code and had to find a suitable time discretisation for this. We first considered the propagation of a kink. A first result is that there is no dependence of the velocity on the angle of the fork even for the full 2D simulation. We therefore introduced a 1D effective partial differential equation to capture the essential features of the 2D propagation. This model incorporates the junction, using the ideas of graph Laplacian \cite{Caputo2013}; its solutions agree well with the 2D solutions. For the kink propagation in a junction we confirm the existence of a critical velocity given approximately by the simple energy conservation argument. Below this velocity the kink gets reflected by the fork. Above it, it passes through the junction and splits into two kinks that propagate in the two different branches. For breathers there are two parameters, $\omega$ the frequency and $v$ the velocity. We mapped the parameter plane $(v,\omega)$ and showed that for
branches of equal width, breathers cross when $v > 1 - \omega$. After the passage through the junction it gives rise to new breathers in the branches. We characterize these new breathers using their energy density and estimate their velocity and frequency. We always observe an up-shift of the frequency and a slight down-shift of the velocity.

The article is organized as follows. In Section II we derive the 1D effective model from the the 2D sine--Gordon equation defined in the fork. In Section III we recall the energies for the kink and the breather and show how they can be used to estimate a critical velocity. Section IV introduces energy conserving discretisations for the finite element 2D problem and the 1D effective equation. Their solutions are compared in section V for both kink and breather initial conditions. Conclusions are presented in Section VI.

\section{Sine--Gordon 2D and 1D effective model}

We consider the 2D sine--Gordon equation
\be\label{eq:2dsg}
  \phi_{tt} - \Delta\phi + \sin\phi = 0,
\ee
on a bounded domain $\Omega \subset \R^2$ with Neuman boundary conditions:
\begin{equation*}
  \nabla\phi\cdot\mathbf{n} = 0,
\end{equation*}
where $\mathbf{n}$ is an exterior normal. The $t$ subscript indicates time derivative and $\Delta$ is the usual Laplacian. This equation conserves the energy
\be\label{eq:en_sg}
  \E = \int_{\Omega} \left[{1\over 2}\phi_t^2 + {1\over 2}|\nabla\phi|^2 
+ (1 - \cos\phi)\right]\,\ud x\,\ud y.
\ee
This can be checked easily by multiplying \eqref{eq:2dsg} by $\phi_t$, integrating over the domain and using the Stokes formula for the spatial operator.

\subsection{1D effective model}

\begin{figure}
  \centering
  \vspace{1em}
\resizebox{9 cm}{8 cm}{\includegraphics{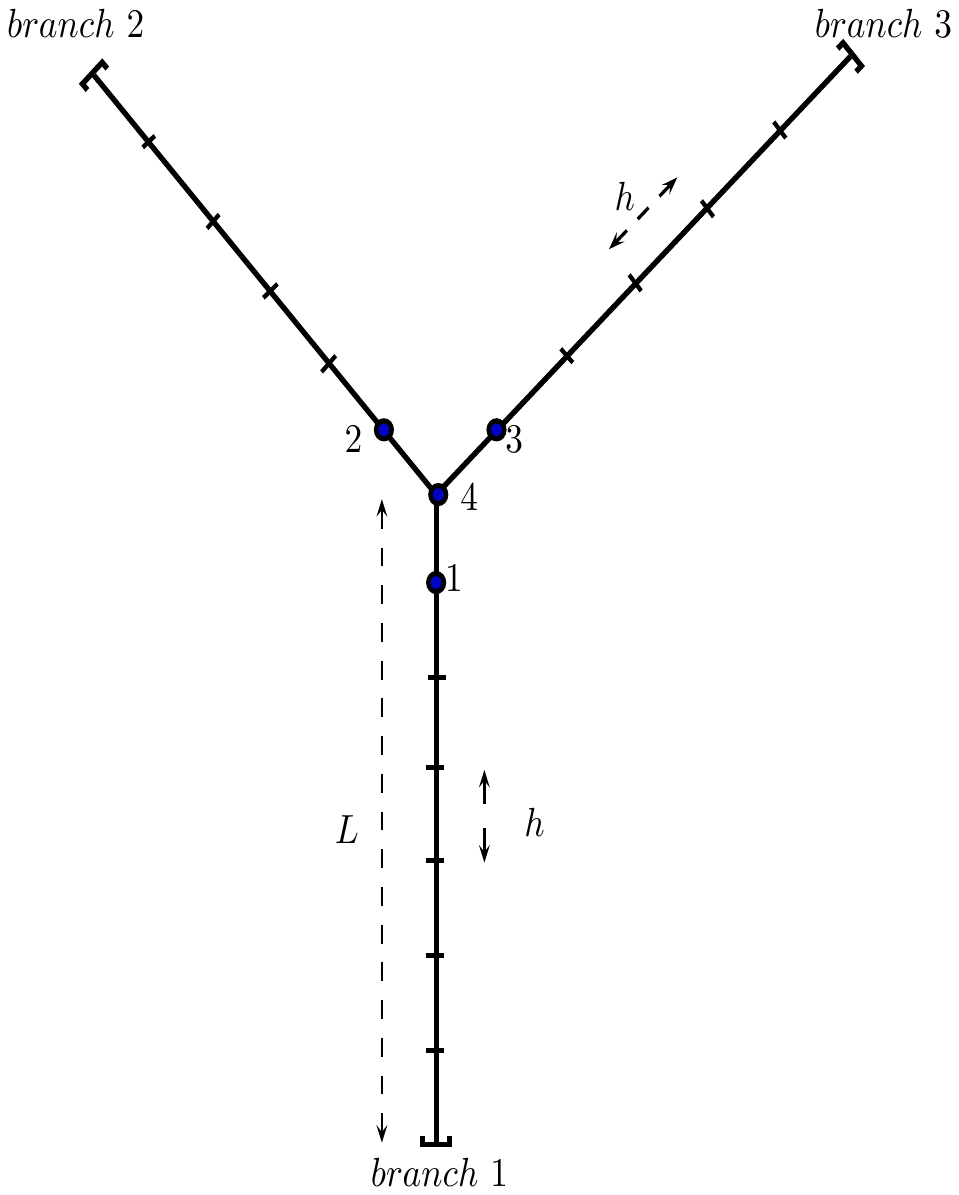}}
\vspace{-2 cm}
  \caption{\small\em Sketch of the tree geometry for the 1D effective model.}
  \label{fig:f2}
\end{figure}

\begin{figure}
  \centering
  \vspace{1em}
  \resizebox{9 cm}{4 cm}{\includegraphics{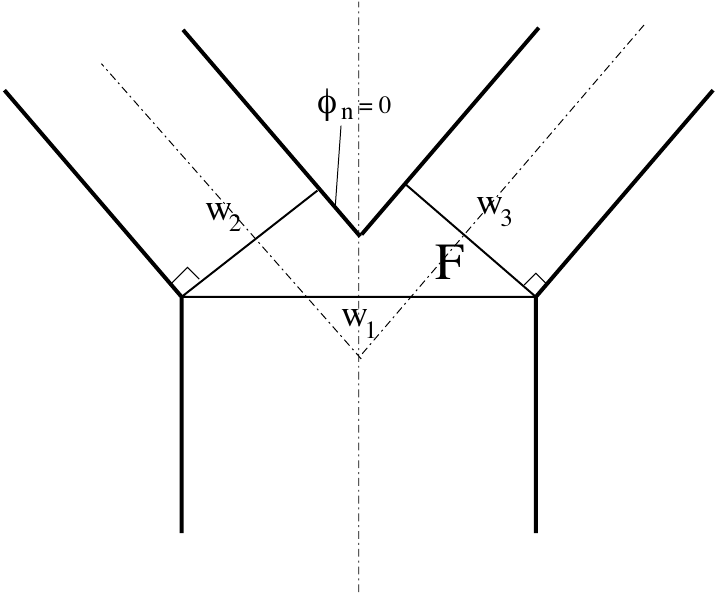}}
  \caption{\small\em The fork region $F$.}
  \label{fig:fork}
\end{figure}

Since the boundary conditions of the 2D problem are homogeneous Neuman it is natural to assume that the solution is uniform in the transverse direction. In other words we keep only the first transverse Fourier mode. Then equation \eqref{eq:2dsg} reduces in each branch to a 1D sine--Gordon equation, 
\be\label{eq:1deff}
  \phi^i_{tt} - \phi^i_{xx} + \sin\phi^i = 0, ~~i=1,2,3 ~,
\ee
where the label $i$ corresponds to the three branches as shown in Fig.~\ref{fig:f2}. These equations are coupled at the apex by two conditions; one is the continuity of $\phi^i$ 
\be\label{cont}
  \phi^1(x=l) = \phi^2(x=0) = \phi^3(x=0)~~,
\ee
and the other is the flux conservation or the Kirchoff law
\be\label{flux}
  w_1 {\phi^1_x} + w_2 {\phi^2_x} + w_3 {\phi^3_x} = 0,
\ee
where ${\phi^i_x}$ is the normal velocity in branch $i$. Let us now briefly justify this flux relation. For that, consider the fork domain $F$ obtained by taking the normals at the different branches as close as possible to the fork as shown in Fig.~\ref{fig:fork}. Integrating the two-dimensional equation \eqref{eq:2dsg} on $F$ yields
\be\label{int_fork}
\int_F (\phi_{tt} + \sin\phi) -\int_{\partial F} \nabla \phi \cdot \mathbf{n}=0,\ee
where $\mathbf{n}$ is the normal to the edge of the domain ${\partial F}$. The second term is equal to the left hand side of \eqref{flux}. The first one is of order $w^2$. In the limit of small width of the branches, $w_i \to 0$ with $w_2/w_1$ and $w_2/w_1$ constant, the first term vanishes while the second one remains.

The numerical scheme used to solve this 1D effective model is described below (see Section~\ref{sec:num}); it is a finite difference approximation. The junction corresponds to the four nodes highlighted on Fig.~\ref{fig:f2}; these are labeled as $1$, $2$, $3$ for the three branches and are connected to the central node $4$. The outer nodes are the last nodes updated by the PDE solver; let us name the value of the solution there $\phi_1$, $\phi_2$, $\phi_3$ for each branch. The value at the central node $\phi_4$ can be computed from the interface conditions \eqref{cont} and \eqref{flux}. Using a forward finite difference approximation for ${\phi^i_x}$ we get from \eqref{flux}
\begin{equation*}
  w_1 (\phi_1 - \phi_4 ) + w_2 (\phi_2 - \phi_4) + w_3 (\phi_3 - \phi_4) = 0,
\end{equation*}
where we have assumed the same space step on the three branches and used the notation $\phi_i \equiv \phi(x_i)$. We have also omitted the $j$ index corresponding to the different branches. We then obtain
\be\label{eq:t4}
  \phi_4 = {w_1\phi_1  + w_2\phi_2 + w_3\phi_3 \over w_1 + w_2 + w_3}. 
\ee

\section{Theoretical considerations}

In 1D the sine--Gordon equation is integrable, see for example \cite{Scott2003}. It has two families of localized exact solutions, the kink
\be\label{eq:kink}
  \phi(x,t) = 4\arctan\left[\exp(\gamma(x - vt))\right],
\ee
and the breather \cite{Dodd1984}
\be\label{eq:breather}
  \phi(x,t) = 4\arctan\left[{\sqrt{1 - \omega^2}\cos(\omega\gamma(t - vx)) 
  \over \omega \cosh(\sqrt{1 - \omega^2}\gamma(x-vt))} \right],
\ee
where the Lorentz factor $\gamma$ is given by
\be\label{eq:gama}
  \gamma = {1 \over \sqrt{1 - v^2}}.
\ee 
Let us consider the kink first. Its energy is
\be\label{eq:en_kink}
  \E_k = 8\gamma.
\ee
The energy of the breather depends also on the frequency. It is given by
\be\label{eq:en_breather}
  \E_b = 16\gamma \sqrt{1 - \omega^2}.
\ee

In two dimensions the equation is not integrable. In addition there is the complication of the boundaries. Therefore the only relations that can be used are conservation laws, and in particular the conservation of energy. When the kink is in branch 1, its energy is $8 w_1 \gamma$ because it is homogeneous in the transverse direction. Similarly in branch 2, it has energy $8 w_2 \gamma_2$. The conservation of energy reads
\be\label{eq:en_junc}
  w_1 {8\over\sqrt{1-v_1^2}} = 2w_2 {8\over\sqrt{1-v_2^2}}.
\ee
This expression gives a critical velocity $v_1$ for which $v_2 = 0$:
\be\label{eq:v_k}
  v_k = \sqrt{1 - \Bigl({w_1 \over 2 w_2}\Bigr)^2}.
\ee
This formula was derived in \cite{Gulevich2006} and tested in a given configuration with success. In the next section we confirm by numerical simulations that this is a good estimate and show its limitations.

A similar argument for the breather yields the following result for the parameters $\{v_1, \omega_1\}$ in the bottom branch and the parameters $\{v_2, \omega_2\}$ in the top branches
\be\label{eq:en_junc_bre}
  {v_1^2 - 1 \over \omega_1^2-1} = \Bigl({w_1 \over 2 w_2}\Bigr)^2 {v_2^2 - 1 \over \omega_2^2-1}.
\ee
This gives a critical velocity $v_1$ for which $v_2 = 0$
\be\label{eq:v_bk}
  v_k = \sqrt{1 - {\omega_1^2 - 1 \over \omega_2^2 - 1}\Bigl({w_1 \over 2 w_2}\Bigr)^2}.
\ee
The practical application of the previous formula is difficult because $\omega_2$ remains unknown. Note however that for small amplitudes, i.e. in the linear limit $\omega_1=\omega_2$ so that we recover \eqref{eq:v_k} for the critical velocity. 

\section{Numerical methods}\label{sec:num}

We now describe the numerical methods used to solve the 2D and the 1D effective problems. We solve \eqref{eq:2dsg} using the finite element method. For that we recall the standard scalar product in $L^2(\Omega)$:
\begin{equation*}
  (\phi, \psi) \equiv \int\limits_{\Omega}\phi\psi\,\ud x\,\ud y.
\end{equation*}

Using this scalar product we project the operator on a test function and use the Green's theorem to integrate the Laplacian \cite{Argyris1991}. For now we did not discretize in time. To do this, we approximate the second derivative in time by a standard three step discretisation. We also average the Laplacian over the current and the following time steps. The final semi-discrete scheme is the following weak formulation
\begin{multline}\label{eq:weak_sg}
  {1 \over \Delta t^2}\left(\phi^{n+1} - 2\phi^n + \phi^{n-1},\psi\right) + \\ {1 \over 2}\left(\nabla (\phi^{n+1} + \phi^n), \nabla\psi\right) + \bigl(\sin\phi^n,\psi\bigr) = 0,
\end{multline}
where $\psi\in L_2(\Omega)$ is the test function, $\Delta t$ is the time-step and $\phi^{n-1}$, $\phi^n$, $\phi^{n+1}$ are respectively the solution at times steps $t_{n-1}$, $t_n$, $t_{n+1}$, where $t_j := j\Delta t$. For the spatial discretization we use a non-structured triangular mesh with $\P_2$ finite element space. The computations are performed using the \texttt{FreeFem++} open-source software \cite{Hecht2012}. The boundary conditions are set to be homogeneous Neuman:
\begin{equation*}
  \nabla\phi\cdot\mathbf{n} = 0,
\end{equation*}
where $\mathbf{n}$ is an exterior normal to the boundary of the domain $\Omega$.

\begin{figure}
  \centerline{\resizebox{9 cm}{6 cm}{\includegraphics{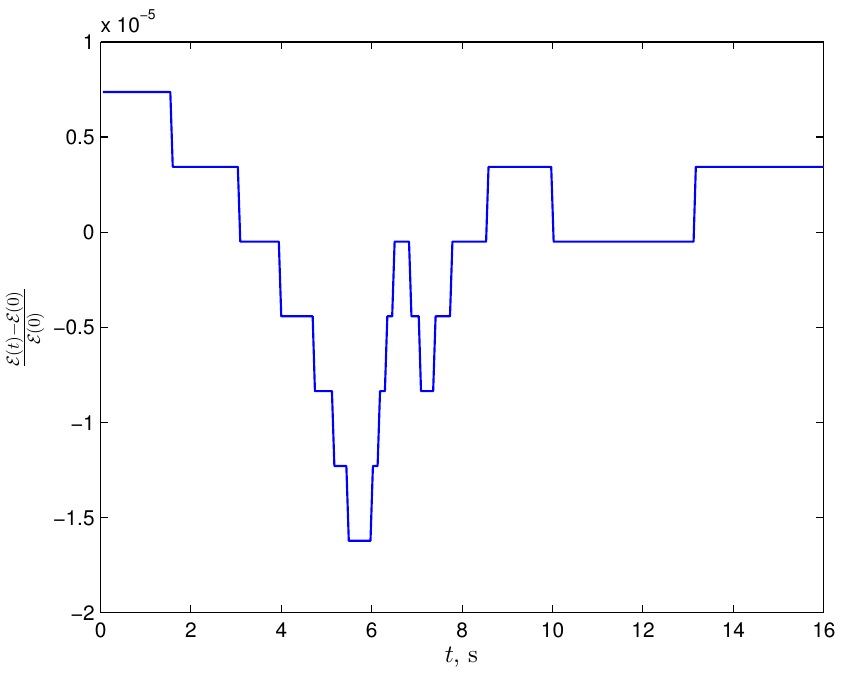}}}
  \caption{\small\em Relative energy $|\E_n - \langle\E_n\rangle|/\langle\E_n\rangle$ for the 2D finite-element solution of a breather propagating in a 2D domain. The symbol $<\cdot>$ denotes the average operator.}
\label{fig:TotalEnergy}
\end{figure}

The total discrete energy is calculated as
\be\label{eq:dis_en}
  \E_n = {1 \over 2} \int \left[\Bigl({\phi^{n+1}- \phi^{n-1}\over 2 \Delta t}\Bigr)^2 + |\nabla\phi^n|^2 - 2(1 - \cos\phi^n)\right]\,\ud x\,\ud y.
\ee
This quantity is conserved up to order $\O(h^4)$, where $h$ is the typical space step. There is no trend in the relative error on the total energy $|\E_n - \E_0|/E_0$ in the course of the computations as shown in Fig. \ref{fig:TotalEnergy}. The time for this plot corresponds to breather of velocity $v = 0.8$ and frequency $\omega = 0.3$ crossing the fork (see Section V). For a kink the error is even smaller. In the numerical simulations presented below we used the mesh with a typical size $\Delta x \approx 0.05$ and the time step $\Delta t = 0.0075$. Because of the implicit nature of the scheme \eqref{eq:weak_sg}, we could take a much bigger time step. However, we preferred to keep it small enough in order to vanish the time discretization error.

The one dimensional effective problem is solved using the finite difference method. The scheme employed reads:
\begin{multline}\label{eq:crank}
  {\phi^{n+1}_{j} + \phi^{n-1}_{j} - 2\phi^n_j \over \Delta t^2} \\ - {1 \over \Delta x^2} \left(\phi^{n}_{j+1} + \phi^{n}_{j-1} - 2\phi^{n}_{j}\right) + \sin\phi^n_j = 0,
\end{multline}
where $n$ and $j$ are the time and space indices correspondingly. Despite the simplicity of the scheme \eqref{eq:crank} it can be shown that in fact, it is a symplectic Euler method derived for the sine--Gordon equation recast in the Hamiltonian form \cite{Leimkuhler2004}. Consequently, the 1D scheme also enjoys good stability and energy conservation properties. Typical values of the space step and the time step are $\Delta x = 0.05$ and $\Delta t = 0.01$.

\section{Numerical results}

\begin{figure}
  \centerline{\resizebox{10 cm}{8 cm}{\includegraphics{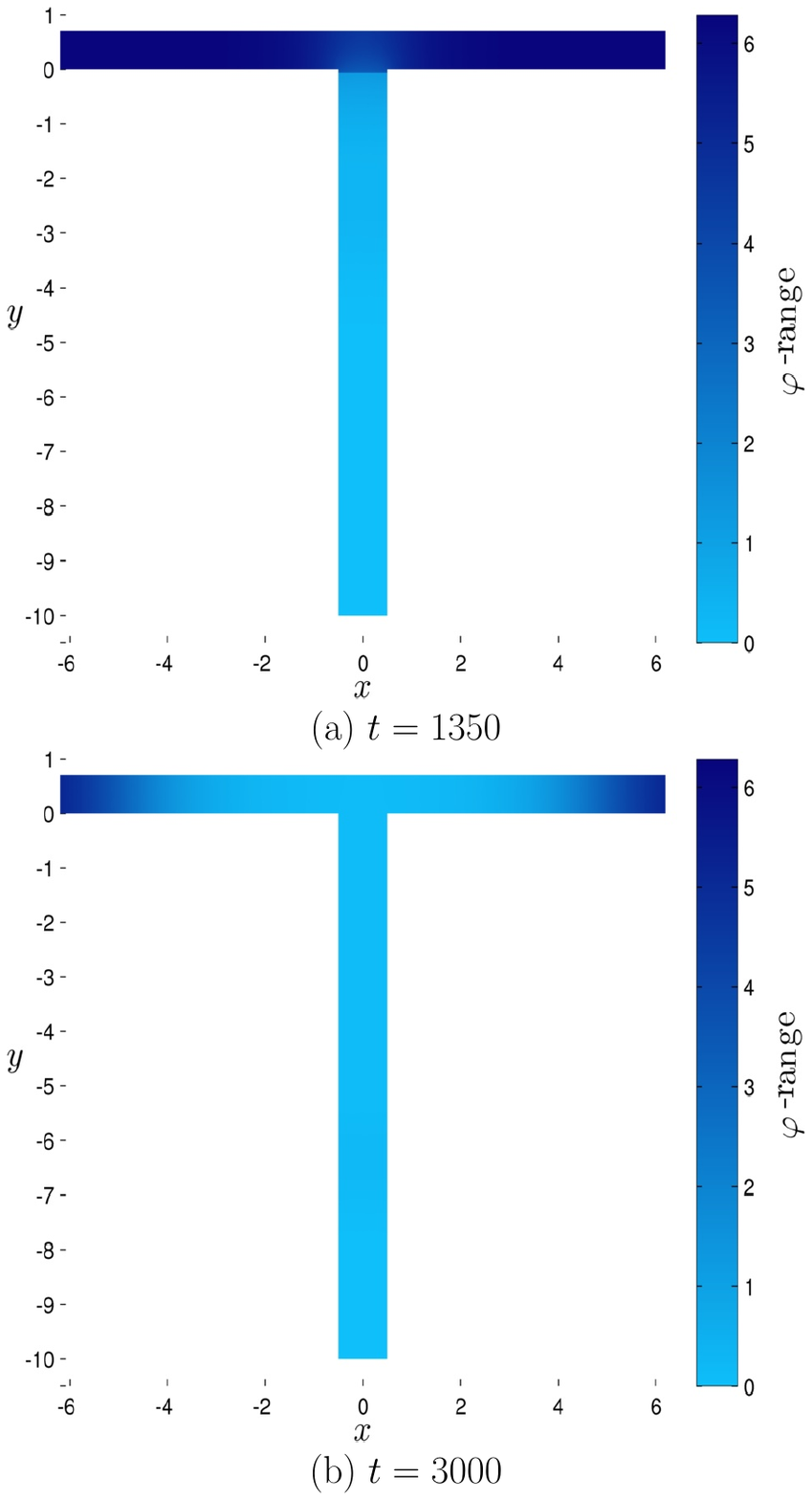}}}
\caption{\small\em Motion in a T--junction. Snapshots of a kink starting in branch 1 with a velocity $v_1 = 0.75$. The values of the time are $t = 1350$ (a) and $t = 3000$ (b) .}
\label{fig:kink_90}
\end{figure}

\begin{figure}
  \centerline{\resizebox{10 cm}{8 cm}{\includegraphics{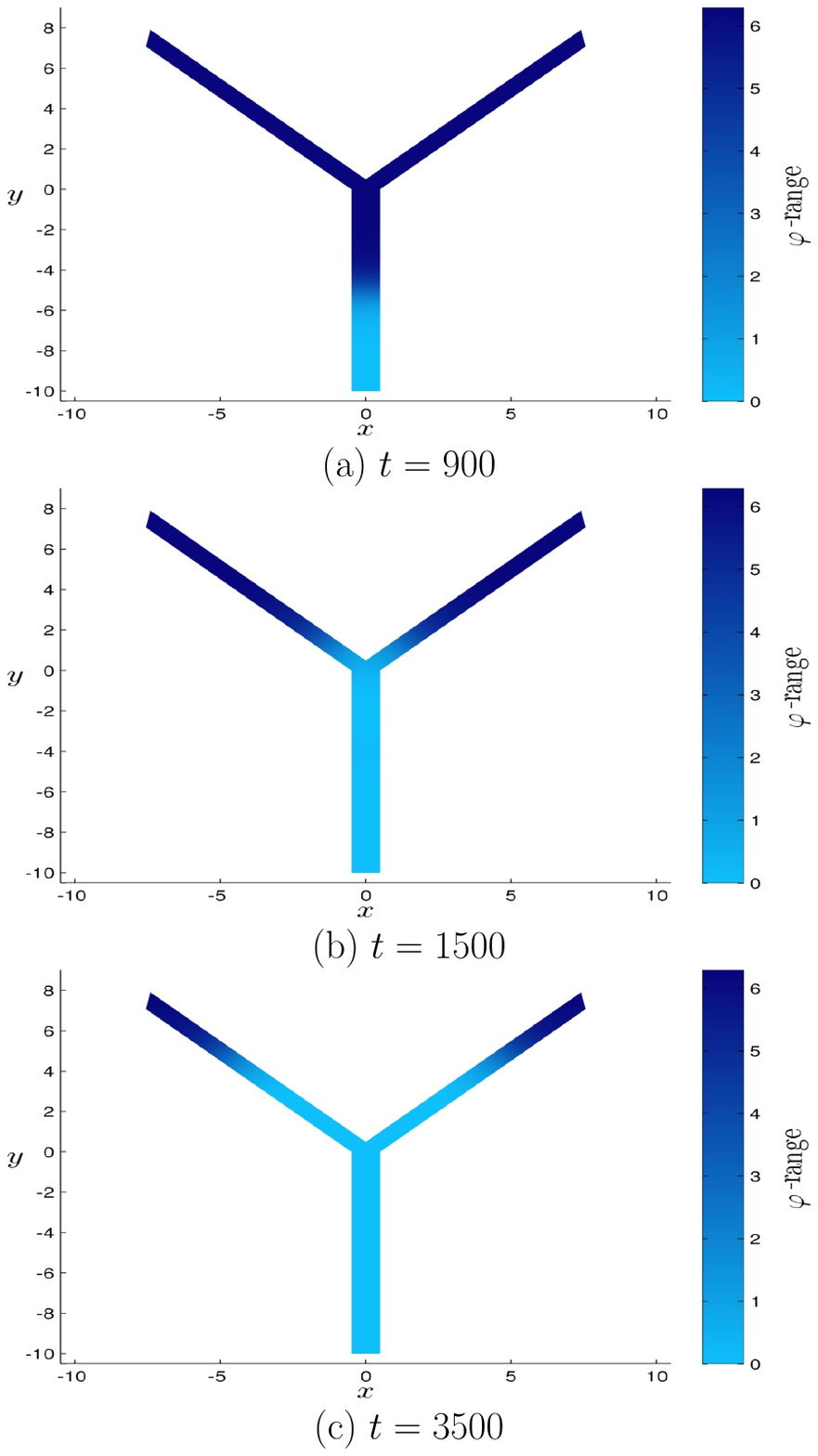}}}
  \caption{\small\em Motion of a kink in a $90^\circ$ Y--junction. Snapshots of a kink starting in branch 1 with a velocity $v_1 = 0.75$. The values of the time are $t = 900$ (a), $t = 1500$ (b) and $t = 3500$ (c).}
  \label{fig:kink_45}
\end{figure}

\begin{figure}
  \centerline{\resizebox{9 cm}{6 cm}{\includegraphics{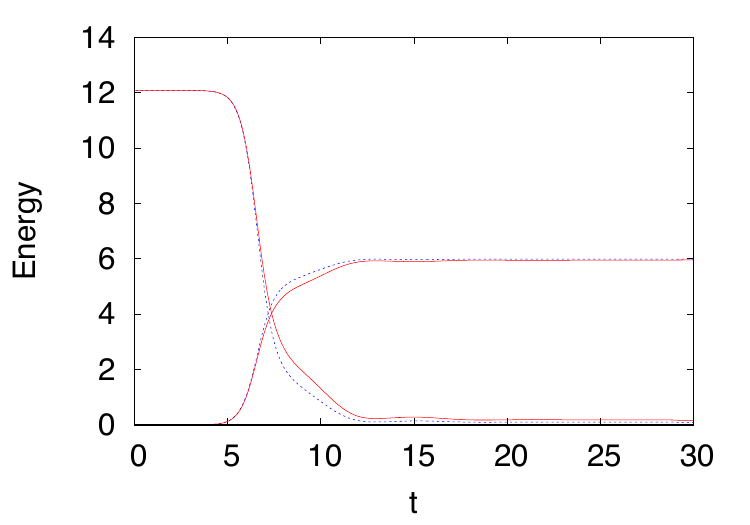}}}
  \caption{\small\em Time evolution of the energy for the kink motion in branch 1 and branch 2 for the T--junction in full line (red online) and for the Y--junction in dashed line (blue online). The parameters are the same as in Figs.~\ref{fig:kink_90} and \ref{fig:kink_45}}.
  \label{fig:ekink90_45}
\end{figure}

We consider the propagation of a kink in Y-- and T--junctions. Our main findings are that the kink gets reflected if it does not have enough energy (velocity). Also the motion depends very weakly on the angle. To illustrate this fact, we show in Fig.~\ref{fig:kink_90} a kink propagating in a T--junction and crossing it. We take the same kink and run it into a Y--junction. This is shown in Fig.~\ref{fig:kink_45}. As can be seen the time intervals for propagation are about the same. This can be seen very clearly when examining the evolution of the energy in the branches 1 (bottom) and 2 (left); this is displayed in Fig.~\ref{fig:ekink90_45}. Note also that a very small amount of energy, typically 5\% of the total energy, is left in branch 1 once the kink has crossed over into branches
2 and 3.

We now compare systematically the 2D solution with the one of the 1D effective equation. This is to validate this approximation. We have conducted a parametric study where we varied $w_1$, $w_2$ and $w_3$ such that $w_2 = w_3 = w_1 + \alpha$ where $w_1 = 1$ and $\alpha = -0.3$, $-0.1$, $0.1$, $0.3$. The results for the critical velocity as well as the estimate \eqref{eq:v_k} are reported in the Table~\ref{tab:energy}. The 2D and 1D models are very close even for $\alpha > 0$. On the other hand the energy estimate is a lower estimate for $\alpha > 0$, The 2D and 1D effective results reveal that the kink crosses the junction but that there are oscillations. The front seems to oscillate and then reshape as it enters more into branches 2 and 3. We do not see this effect when $\alpha \le 0$. Despite this, the values are all within a 10\% interval of error.

\begin{table}
\centering
\begin{tabular}{|c|c|c|c|}
   \hline\hline
   $\alpha$     &  2D $v_c$   &  1D $v_c$  &  $v_k$ from \eqref{eq:v_k} \\ \hline
   \hline\hline
   0.3          &  0.98       &  0.99      &   0.92 \\ \hline
   0.1          & 0.965       &  0.955     &   0.89 \\ \hline
   0            & 0.92        &  0.94      &   0.86 \\ \hline
   -0.1         & 0.885       &  0.85      &   0.83 \\ \hline
   -0.3         & 0.73        &  0.71      &   0.7 \\
   \hline\hline
\end{tabular}
\caption{\small\em Critical velocities for the 2d model, the 1d effective model and the energy estimate as a function of $\alpha$. The widths of the branches are $w_1 = 1$, $w_2 = w_3 = w_1 + \alpha$.}
\label{tab:energy}
\end{table}

\begin{figure}
  \centerline{\resizebox{10 cm}{7 cm}{\includegraphics{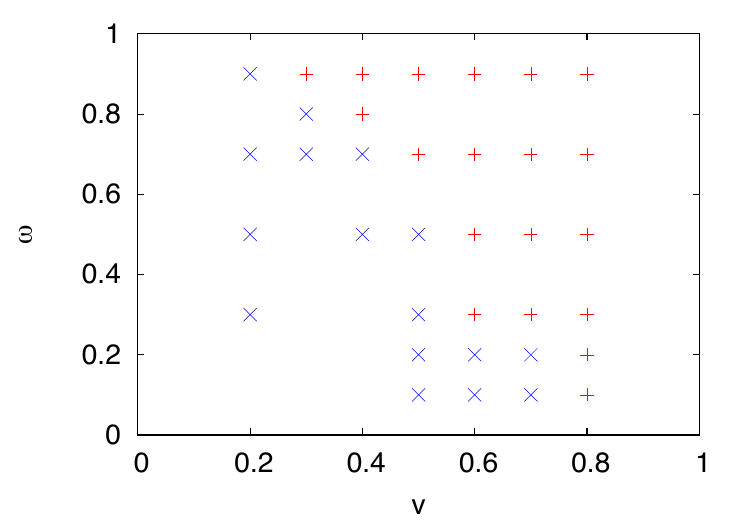}}}
  \caption{\small\em Parameter space $(v,\omega)$ for the crossing of breathers obtained from the 1D effective model. The small $+$ symbol (red online) corresponds to the breather crossing while the $\times$ symbol (blue online) corresponds to the breather being reflected by the junction.}
  \label{fig:pass}
\end{figure}

For the breather, things are more complicated because of the additional parameter, the frequency. Fig. \ref{fig:pass} shows the parameter space $(v, \omega)$. The crossing (resp. reflection) of the breather is indicated by the $+$ (resp. $\times$) sign. One can see that for large enough velocities, the breather crosses independently of its frequency. On the other hand, for frequencies close to one, the breather crosses even for small velocities. This situation is close to the linear case for which we expect always some energy transfer to the other branch \cite{bp09}.

\begin{figure}
  \centerline{\resizebox{9 cm}{6 cm}{\includegraphics{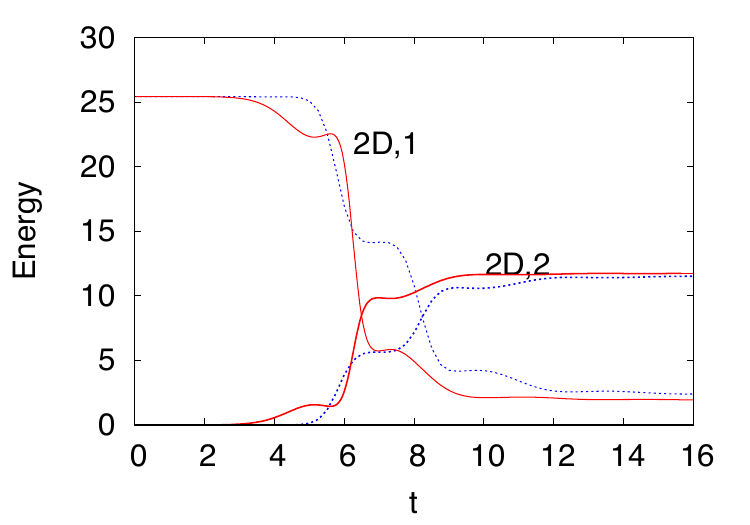}}}
  \caption{\small\em Time evolution of the energy in branches 1 and 2 for a breather initial condition for the 2d partial differential equation in full line (red online) and the 1d effective model in dashed line (blue online). The parameters are $w_i = 1$, $i=1,2,3$, $v_1 = 0.8$, $\omega_1 = 0.3$, $x_0 = 10$.}
  \label{fig:en_brea}
\end{figure}

There is always a small reflection from the fork. For example, we show the time evolution of the energy of a breather in Fig.~\ref{fig:en_brea}. Notice how the energy in branch 1 does not drop to 0 as for the kink. There is a remainder.

\begin{figure}
  \centerline{ \resizebox{10 cm}{8 cm}{\includegraphics{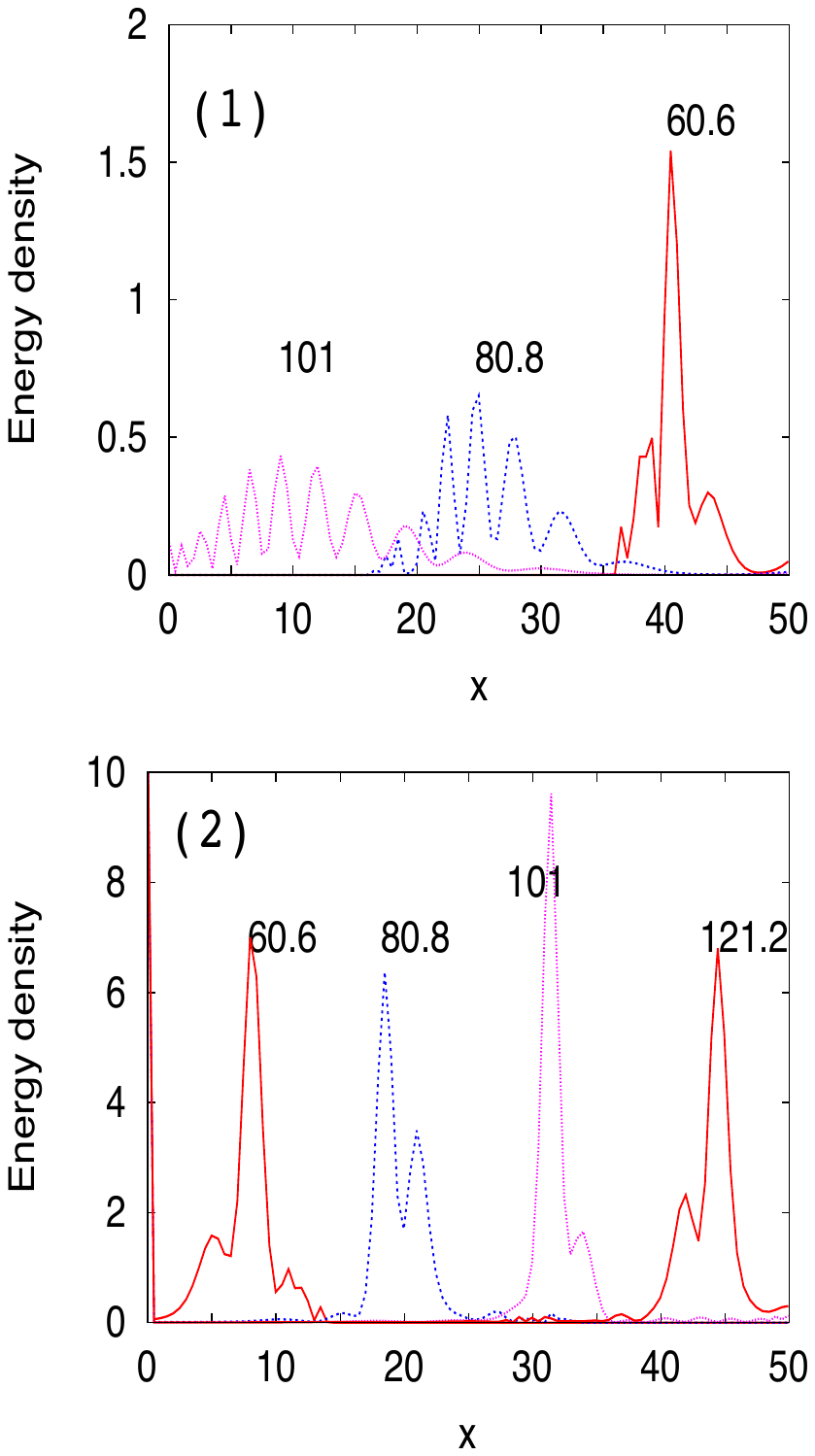} }}
  \vskip -1 cm
  \caption{\small\em Snapshots of the energy density of a breather at three different times in branches 1 (top) and 2 (bottom). The parameters are $w_i = 1$, $i = 1, 2, 3$, $v_1 = 0.8$, $\omega_1 = 0.3$. The initial position of the breather is $x_0 = 10$.}
\label{fig:ee1}
\end{figure}

To characterize the breathers in the other branches is difficult because the wave oscillates. We found that plotting the energy density 
\be\label{eq:density}
  d\E = {1\over 2}\phi_t^2 + {1\over 2}\phi_x^2 + 1 - \cos\phi  ,
\ee
gives a good indication of the position of the breather. Let us analyze in more details the specific configuration where a breather of speed $v = 0.8$ and frequency $\omega = 0.3$ crosses the junction. Fig.~\ref{fig:ee1} shows the energy density for three different times in the branch 1 (left panel) and in the branch 2 (right panel) after the breather has passed the junction. Then the energies in branch 1 and branch 2 are respectively $\E_1 = 2.16 $ and $\E_2 = 13.23$. The velocities estimated by a least square fit on the center of mass of the breather density are respectively $v_1 = -0.75$ and $v_2 = 0.6$. They are lower than the initial velocity to accommodate for the crossing of the breather. The frequencies of the breathers in branches 1 and 2 can be estimated, they are respectively $\omega_1 = 0.996$ and $\omega_2 = 0.75$. All these parameters are very different from the initial breather parameter making the scattering of a breather much more complex than the one of a kink.

Using the parameters above we can plot the fitted breathers and compare them with the numerical solution. Fig.~\ref{fig:ee1} shows in the top panel, branch 1 before the breather crosses. There the analytical solution matches perfectly the numerical one. The middle and bottom panels show respectively branch 2 and branch 1 after the crossing. Here the agreement is not as good but remains
quite acceptable.

\begin{figure}
  \centering
  \centerline{\resizebox{9 cm}{8 cm}{\includegraphics{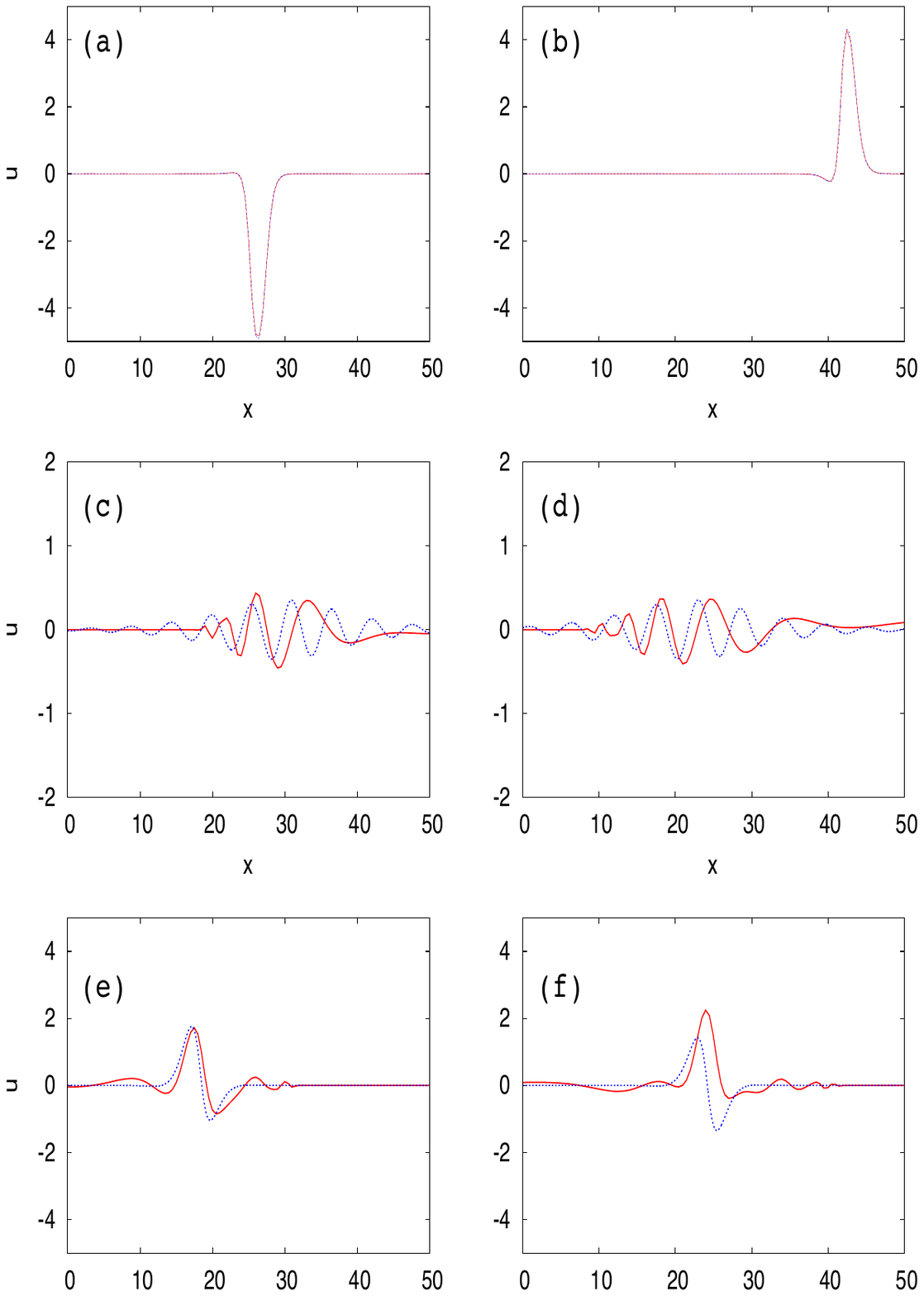}}}
  \caption{\small\em Snapshots of the breather analytical solution with the fitted parameters, before and after the crossing in branch 1 (panels (a, b, c, d) and after the crossing in branch 2 (panels (e, f)). The corresponding times are $t = 20.2$, $40.4$, $80.8$, $90.9$ for panels (a, b, c, d) and $t = 80.8$, $90.9$ for panels (e) and (f).}
  \label{fig:brea}
\end{figure}

To conclude this study we examine systematically the influence of the breather frequency on its crossing. We took $v_1 = 0.8$ and chose $\omega_1 = 0.3$, $0.5$, $0.7$ and $0.9$. The results are reported in the Table~\ref{tab:breather}.

\begin{table}
\centering
\begin{tabular}{|c|c|c|c|c|c|c|c|}
  \hline
branch              &  1   &  1-return  &  2    & ~~  & 1   &  1-return  &  2\\ 
index $i$           &      &            &       & ~~  &      &            &         \\ \hline
  $\omega_i$        &  0.3 & 0.99       &  0.79 & ~~  & 0.5 &  0.99      & 0.87 \\ \hline
  $v_i$             & 0.8  &  0.8       &  0.56 &  ~~ &0.8  &  0.8        &   0.65 \\ \hline
  energy {\cal E}   & 25.42& 2.1        &11.66  &  ~~ & 23.07&   2.12     &  10.48  \\ \hline
                    &      &            &       &  ~~ &      &            &         \\
                    &      &            &       &  ~~ &      &            &         \\
  $\omega_i$        &  0.7 & 0.998      &  0.93 & ~~  &  0.9 & 0.999      &  0.98 \\ \hline
  $v_i$             & 0.8  &  0.85      &  0.73 & ~~  &  0.8 &  0.85      &   0.8 \\ \hline
  energy {\cal E}   & 19.03& 1.91       &  8.57 & ~~  & 11.61&   1.23     &  5.192  \\ \hline
\end{tabular}
\caption{\small\em Velocities and frequencies for the crossing of a breather
of initial velocity $v_1=0.8$ and different frequencies $\omega_1 = 0.3$, $0.5$ (top rows) and $\omega_1 = 0.7$, $0.9$ (bottom rows). The columns indicate the branches, 1, ``1-return'' and 2. The label ``1-return'' corresponds to branch 1 after the collision.}
\label{tab:breather}
\end{table}

\section{Conclusion}

We analyzed numerically and theoretically how a 2D sine--Gordon kink or breather crosses a Y-- or T--junction. The similarities between the energies in the different branches for both cases shows that the angle of the junction plays almost no role in the dynamics.

This suggested to introduce a 1D effective model where at the junction, we satisfy continuity of the solution and a jump condition for the gradient given by the conservation of flux. The solutions of this effective model accurately reproduce the 2D solutions.

The parameters for the kink to cross obey the simple relation obtained from the conservation of energy. There is a critical velocity below which no crossing is possible.

On the contrary the breather crossing is more complex. There are two parameters: the velocity $v$ and frequency $\omega$. For equal widths of the branches, we observe that crossing happens when $v > 1 - \omega$. Then the breather gives rise to other breathers in the two upper branches that we characterize using the energy density and the value of the energy. These new breathing solutions propagate slower than the initial condition and it is also up-shifted in frequency. We always observe a small reflexion at the crossing into the first branch.

This study can be extended by considering more branches. Another interesting extension would be to add a source at the junction, enabling to control the crossing. It would be useful to understand how this study can be generalized to another application, like the reflexion of shallow water waves.
\bigskip

\begin{acknowledgments}
D.D. acknowledges the support from ERC under the research project ERC-2011-AdG 290562-MULTIWAVE and thanks INSA de Rouen for its hospitality during his visit in December 2012. The authors thank D.~Mitsotakis and G.~Sadaka for useful discussions on the finite element numerical method. The authors acknowledge the Centre de Ressources Informatiques de Haute Normandie where most of the calculations were done.
\end{acknowledgments}

\end{document}